\documentclass[twocolumn,prl,showpacs,superscriptaddress]{revtex4-1}
\usepackage[dvips]{graphicx}%
\usepackage{dcolumn}
\usepackage{amsmath}
\begin{document}

\title{Bulk-like viscosity and shear thinning during dynamic compression of a nanoconfined liquid}
\author{Shah H. Khan}
\affiliation{Institute of Physics and Electronics, University of Peshawar, Peshawar, 25120, Pakistan}
\author{Peter M. Hoffmann}
\email{hoffmann@wayne.edu}
\affiliation{Department of Physics and Astronomy, Wayne State University, Detroit, MI 48201, USA}
\date{\today}

\begin{abstract}
The viscosity of liquids under nanoconfinement remains controversial. Reports range from spontaneous solidification to no change in the viscosity at all. Here, we present thorough measurements with a small-amplitude linear atomic force microscopy technique and careful consideration of the confinement geometry, to show that in a weakly interacting liquid, average viscosity remains bulk like, except for strong shear thinning once the liquid is confined to less than four molecular layers. Overlaid over this bulk-like viscous behavior are stiffness and damping oscillations, indicating non-continuum behavior, as well as an elastic response when the liquid is allowed to order in the confinement gap.

\end{abstract}
\maketitle

Nanoconfined liquids play an important role in nanotechnology\cite{Dai10,Boc10}, tribology \cite{Bhu95}, geology \cite{Che12}, biology \cite{Fin96} and medical technology\cite{Pra12}, but their properties remain controversial \cite{Kle98,Ant01, Dem01,Rav02,Bec03,Li07,Mat08,Kag08, Hof09,Li08,Pat06,Kha10}. Although it has been widely accepted that nanoconfined liquids are different from the bulk, their often surprising viscoelastic behavior is not yet completely understood. According to some experimental groups, liquids pass through a spontaneous phase transformation from liquid to solid below a certain threshold confinement \cite{Kle98}. Other groups see an enhancement in viscosity or relaxation time only \cite{Dem01}, or see very little change in viscosity upon confinement\cite{deB10}. The behavior may also be different for different liquids and depend on the dynamic rate of confinement \cite{Zhu03,Pat06,Li08,Kha10,Bur10}.

Atomic force microscopy (AFM) or surface force apparatus (SFA) measurements of the mechanics of nanoconfined liquids either measure a damping coefficient under compression, involving squeeze-out\cite{Osh98}, or employ shear measurements\cite{Ant01,Mat08}. While the latter leads to a more straight-forward measure of effective viscosity, squeeze out measurements are often easier to perform and can be combined with measurements of the compressive elasticity of the nanoconfined liquid. Moreover, some recent reports suggest that the behavior of the liquid may be anisotropic, i.e. viscous behavior may be different dependent on the load direction \cite{Bur10}.

A problem with comparing measurements is assessing the influence of confinement size. In order to normalize the compression force, the Derjaguin approximation is commonly used\cite{Der34}. For a sphere-on-flat, this approximation states that the quantity $F/R$ is independent of confinement geometry, where $F$ is the force and $R$ is the radius of the sphere, which in AFM is the cantilever tip. The Derjaguin approximation holds over many orders of magnitude, and allows for the comparison of SFA and AFM measurements in a variety of geometries. For a sphere-on-flat, the interfacial energy per unit area can be directly related to the interfacial force, as we have $w = dW/dA = -1/(2 \pi R) dW/dh = F/2 \pi R$, where  $h$ is the vertical distance along the sphere, measured from the substrate to the sphere surface. Since the interfacial energy is assumed to be independent of radius, $F/R$ should be a the same for any radius sphere approaching a flat substrate. We use a small amplitude dynamic technique, which does not measure force, but rather the local force gradient or interaction stiffness (where $k_{int} = -dF/dz$) and the squeeze damping, $C$, allowing determination of the effective viscosity. For the interaction stiffness, we find  $k = -dF/dz = d^2 W/dh^2 = -d/dh \left( 2 \pi R dW/dA \right) = - 2 \pi R  dw/dh$. Assuming that the gradient of the interfacial energy is a function of $h$ only, the interaction stiffness $k$ should also scale linearly with the radius of the sphere. Note that typically $dw/dh<0$, and $k$ is positive.

The correct expression for the squeeze damping term can be found by classical lubrication theory from Reynolds equation, which, in the axisymmetric case reads\cite{Wil05}:
\begin{equation}\label{Reynolds}
\frac{1}{r} \frac{\partial}{\partial r} \left( \frac{r h^3}{\eta} \frac{\partial p}{\partial r} \right) = - 12 \frac{\partial h}{\partial t}
\end{equation}
For a sphere with $h(r) = h_0 + R - \sqrt{R^2-r^2}$, constant $\partial h / \partial t = v$ and constant $\eta$, \eqref{Reynolds} can be integrated twice to give
\begin{equation}\label{DampForce}
F= C v = \frac{6 \pi \eta v R^2}{h_0} \left[1+3.5 \frac{h_0}{R} + 3 \frac{h_0}{R} \ln{\frac{h_0}{R}} \right]
\end{equation}
Typically, only the leading term is used, as the correction terms become very small for $h_0\ll R$. However, equation \eqref{Reynolds} can also be integrated for an arbitrary axisymmetric shape of the tip  (for non-symmetric tips, the more general two-dimensional form of the Reynolds equation must be solved, which is significantly more involved) and for non-constant viscosities.

Measurements were performed on nanoconfined tetrakis(2-ethylhexoxy)silane (TEHOS), an inert liquid, which has been well characterized through X-ray\cite{Yu99}, fluorescence correlation \cite{Pat07} and AFM measurements\cite{Mat08}. Stiffness and damping were determined with a sub-resonance dynamic AFM technique, by measuring amplitude and phase of the approaching tips at amplitudes of $<0.6$ nm and an approach speed of 1 nm/s. The AFM cantilevers were completely immersed in the liquid during the measurement avoiding any surface tension effects. A home-built, fiber-interferometric AFM was used for the measurements\cite{Pat05}. Cantilevers, substrates and the measured liquid were thoroughly cleaned prior to measurements. We used four different cantilevers, with stiffnesses $1.0, 2.4, 35$ and $55 \textrm{N/m}$, with nominal tip sizes of $0.5, 1.4, 4.7$ and $18  \mu\textrm{m} $, respectively, in order to explore the effect of tip size on the stiffness and damping of nanoconfined TEHOS.

\begin{figure}\centerline{\includegraphics[width=83mm, clip,
keepaspectratio]{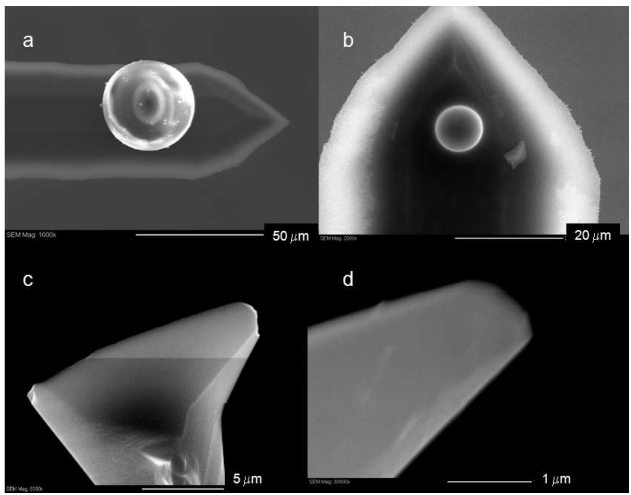}} \caption{Scanning Electron images of the four probes used in this study.}\label{tips}
\end{figure}

\begin{figure}\centerline{\includegraphics[width=83mm, clip,
keepaspectratio]{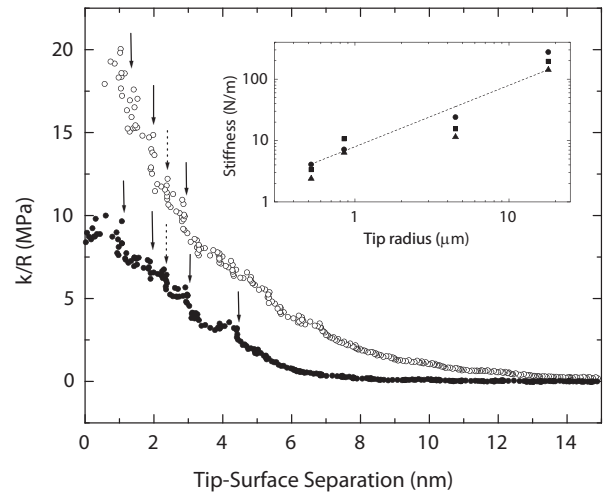}} \caption{Reduced stiffness $k/R$ for the largest $R=18 \mu\textrm{m}$ and the smallest tip radius $R=0.5 \mu\textrm{m}$. Arrows indicate layering of the liquid, with a temporary pinning of the tip and local peaks in the stiffness. Unlike our previous reports, we were able to extract the actual tip position  and therefore could plot these data against the actual tip-surface distance. Dashed arrow shows an intermediate peak apparent in both data sets. Inset: Log-log plot of stiffness $k$ versus tip radius $R$ for all four tips. Dashed line has slope of one, indicating a linear relationship between $k$ and $R$. Triangles: $h=3$ nm, Circles: $h=2$ nm, Squares: $h=1$ nm.}\label{Stiffness}
\end{figure}

Figure \ref{tips} shows scanning electron microscopy (SEM) images of the four tips used in this study. Such imaging is limited, but allows us to approximate the tip shape sufficiently to get close matches between the measured and calculated damping curves. As seen, the radius of curvature $R_c$ of the various tips is not necessarily the same as the nominal radius of the tip end, $R$. This is due to flattening of the tip during the initial measurement stages. This flattening has the advantage that small asperities are eliminated, and unlike other reported measurements \cite{Lim08}, our measurements reproduce the expected Derjaguin dependence of the stiffness on the radius quite well (Figure \ref{Stiffness}). We found that the stiffness is linearly dependent on $R$ and \emph{not} on the radius of curvature of the tip, $R_c$. This is not too surprising, as the Derjaguin approximation relates to the contact radius, which is mostly determined by actual tip size.

\begin{figure}\centerline{\includegraphics[width=83mm, clip,
keepaspectratio]{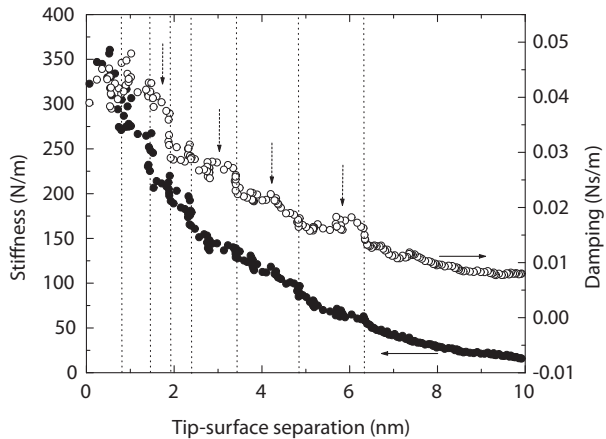}} \caption{Stiffness and damping for the largest ($R=18 \mu\textrm{m}$) probe. Dotted lines indicate layering of the liquid. Arrows point out increased damping between the layers.}\label{StiffandDamp}
\end{figure}

Figure \ref{StiffandDamp} shows stiffness and damping for the largest probe. Stiffness and damping reflect molecular layering of the liquid, exhibiting regular peaks roughly commensurate with molecular size. It should be noted that seeing these peaks with a colloidal probe tip is challenging, as any tip roughness will destroy average ordering in the confined liquid. To our knowledge, this is the first time that stiffness and damping oscillations in a nanoconfined liquid have been seen using a large colloidal AFM probe. The stiffness peaks are at locations where the tip remains stationary, i.e. where the tip is pushing against an integer number of molecular layers. By contrast, the damping peaks preferentially line up \emph{between} the stationary points (dashed arrows), indicating that enhanced damping may result from rearranging molecules as the liquid transitions from $N$ to $N-1$ layers between tip and substrate. However, as shown previously, the fact that stiffness and damping peaks are out of phase with each other is not universal, but depends on the squeeze rate \cite{Pat06,Kha10}. In TEHOS, we almost exclusively observe the damping and stiffness to be out of phase, indicating an elastic-like behavior of the ordered molecular layers of the liquid.

\begin{figure}\centerline{\includegraphics[width=83mm, clip,
keepaspectratio]{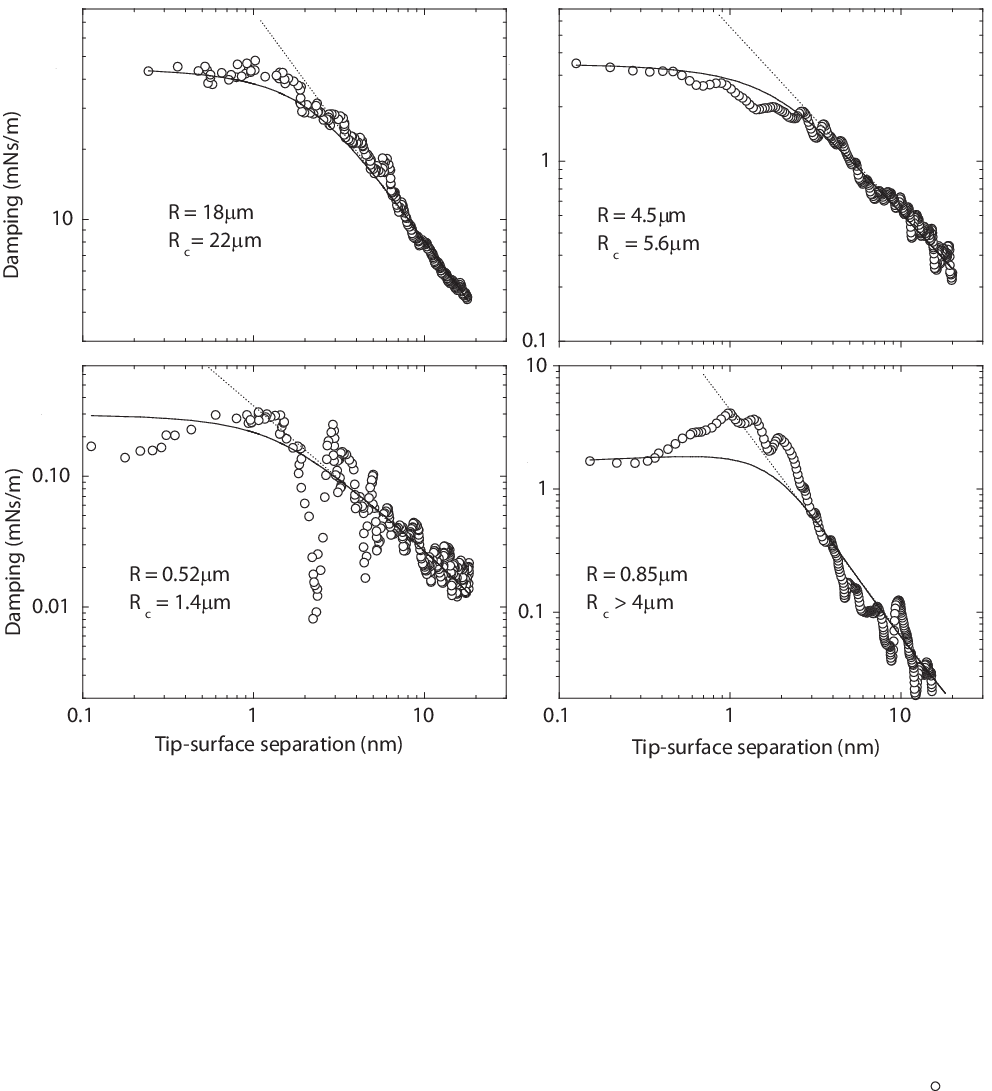}} \caption{Log-log plots of the damping coefficient of the confined liquid for all four tips. Tip parameters are given in the respective figures. Dashed line indicates slope without assuming any shear thinning. This slope is typically equal to -1, as expected for a spherical tip, but deviations can be observed if significant flattening is present.}\label{damping}
\end{figure}

Figure \ref{damping} shows the damping curves measured with the four probes on a log-log plot. Generally, the damping is proportional to $h_0^{-1}$ as expected, although there are deviations. At small separations of a few nanometers, and therefore a few molecular layers, damping saturates. This saturation of the damping can have two origins. Elastohydrodynamic effects can lead to a deformation of the substrate by the localized liquid pressure above it\cite{Ler11}. However, if we calculate the critical liquid gap distance where such effects should be seen, it is of order $10^{-12}$ m, which is much too low to account for the saturation seen at a few nanometers in these measurements.

Alternatively, a saturation in the damping may suggest a reduction of the effective viscosity at close approach. Since at close approach the squeeze-out leads to increased shear in the liquid, this suggests shear thinning. This is consistent with previous reports on shear thinning effects in nanoconfined octa-methyl cyclotetrasiloxane (OMCTS) \cite{Hu91} and in water\cite{Li08}. Shear thinning can be described by the Carreau model\cite{Bir68}, where the effective viscosity is given by
\begin{equation}\label{Carreau}
\eta = \frac{\eta_0}{1+(\lambda\dot{\gamma})^{1-s}}
\end{equation}
where $\lambda$ is the characteristic time scale at which shear thinning becomes observable, and $s$ is the so-called shear thinning exponent (for shear thinning, $s<1$).The shear rate can be estimated by calculating the volumetric flux per unit area as the tip squeezes out the liquid. We find $j = \frac{1}{A}\frac{dV}{dt} = \frac{\pi r^2 v}{2 \pi r h} = \frac{rv}{2h}$.  The average shear rate at an axial distance $r$ is then the flux divided by film thickness, or $\dot{\gamma} = \frac{rv}{2h^2}$. This can be substituted into equations \eqref{Reynolds} and \eqref{Carreau}which can then be numerically integrated for any tip shape $h = h(r)$.

We estimated the tip shape from the SEM images, from which a reasonable range of possible radii of curvature can be obtained. Within this range, we adjusted parameters to obtain the best fit to the measured data. In this sense the tip shape is consistently determined by both imaging and fitting to the obtained damping curve. It should be noted that at large distances, before the damping saturates, $s$ and $\lambda$ have practically no influence on the fit. Thus the fit, together with the SEM imaging, determines the effective tip geometry quite well. We also kept the fitting parameters $\lambda$ and $s$ as close as possible across all measurements. We found that a value of $s=-0.7$ gives best fits for all tip geometries used, while $\lambda$ varied in the relatively narrow range of 90-180 ms. For all fits, we used a viscosity of 0.009 Pa s, which is the expected bulk viscosity of TEHOS at 293 K \cite{Abb61}. This universality of our fits to a variety of tip geometries gives us confidence that no viscosity enhancement was observed in these measurements. The saturation of damping at close approach suggests shear thinning, which is consistent with previous reports \cite{Hu91, Li08}. It should be noted that the fact that shear thinning is observed when the closest tip-surface distance is just a few nanometers does not mean that shear thinning occurs at this separation. The tip is curved, and therefore there is a range of liquid thicknesses at different locations along the radius of the tip. Moreover, maximum shear flow does not occur at the tip center.

In conclusion, we have shown that in a simple, weakly interacting liquid, the average viscosity measured by squeeze-out is bulk-like even for molecularly thin liquid layers. However, at moderate shear rates and very close to the surface, shear thinning is observed. Overlaid on the average bulk-like/shear thinning behavior are oscillations in viscosity commensurate with molecular layering of the liquid. The fact that damping and stiffness peaks are out-of-phase with respect to molecular layering indicates dynamic effects, which lead to an elastic response when the liquid is molecular ordered between the two confining surfaces \cite{Pat06,Kha10}. Thus the behavior of nanoconfined liquids can be surprisingly rich, exhibiting average continuum behavior, down to molecular scale, which is however modulated by discreteness of the system, once it is confined to less than 4-5 molecular layers. This is also reminiscent of plastically deformed solid systems, which can show combine discrete jumps upon deformation, superimposed on an overall continuum-like response \cite{Hof01}.

\begin{acknowledgments}
P.\ M.\ H.\ would like to acknowledge funding through the National Science Foundation, grant DMR-0804283.
\end{acknowledgments}

\bibliography{TEHOS}

\begin{thebibliography}{33}%
\makeatletter
\providecommand \@ifxundefined [1]{%
 \@ifx{#1\undefined}
}%
\providecommand \@ifnum [1]{%
 \ifnum #1\expandafter \@firstoftwo
 \else \expandafter \@secondoftwo
 \fi
}%
\providecommand \@ifx [1]{%
 \ifx #1\expandafter \@firstoftwo
 \else \expandafter \@secondoftwo
 \fi
}%
\providecommand \natexlab [1]{#1}%
\providecommand \enquote  [1]{``#1''}%
\providecommand \bibnamefont  [1]{#1}%
\providecommand \bibfnamefont [1]{#1}%
\providecommand \citenamefont [1]{#1}%
\providecommand \href@noop [0]{\@secondoftwo}%
\providecommand \href [0]{\begingroup \@sanitize@url \@href}%
\providecommand \@href[1]{\@@startlink{#1}\@@href}%
\providecommand \@@href[1]{\endgroup#1\@@endlink}%
\providecommand \@sanitize@url [0]{\catcode `\\12\catcode `\$12\catcode
  `\&12\catcode `\#12\catcode `\^12\catcode `\_12\catcode `\%12\relax}%
\providecommand \@@startlink[1]{}%
\providecommand \@@endlink[0]{}%
\providecommand \url  [0]{\begingroup\@sanitize@url \@url }%
\providecommand \@url [1]{\endgroup\@href {#1}{\urlprefix }}%
\providecommand \urlprefix  [0]{URL }%
\providecommand \Eprint [0]{\href }%
\providecommand \doibase [0]{http://dx.doi.org/}%
\providecommand \selectlanguage [0]{\@gobble}%
\providecommand \bibinfo  [0]{\@secondoftwo}%
\providecommand \bibfield  [0]{\@secondoftwo}%
\providecommand \translation [1]{[#1]}%
\providecommand \BibitemOpen [0]{}%
\providecommand \bibitemStop [0]{}%
\providecommand \bibitemNoStop [0]{.\EOS\space}%
\providecommand \EOS [0]{\spacefactor3000\relax}%
\providecommand \BibitemShut  [1]{\csname bibitem#1\endcsname}%
\let\auto@bib@innerbib\@empty
\bibitem [{\citenamefont {Daiguji}(2010)}]{Dai10}%
  \BibitemOpen
  \bibfield  {author} {\bibinfo {author} {\bibfnamefont {H.}~\bibnamefont
  {Daiguji}},\ }\href@noop {} {\bibfield  {journal} {\bibinfo  {journal}
  {Nature Nanotechnology}\ }\textbf {\bibinfo {volume} {5}},\ \bibinfo {pages}
  {831} (\bibinfo {year} {2010})}\BibitemShut {NoStop}%
\bibitem [{\citenamefont {Bocquet}\ and\ \citenamefont
  {Charlaix}(2010)}]{Boc10}%
  \BibitemOpen
  \bibfield  {author} {\bibinfo {author} {\bibfnamefont {L.}~\bibnamefont
  {Bocquet}}\ and\ \bibinfo {author} {\bibfnamefont {E.}~\bibnamefont
  {Charlaix}},\ }\href@noop {} {\bibfield  {journal} {\bibinfo  {journal}
  {Chemical Society Reviews}\ }\textbf {\bibinfo {volume} {39}},\ \bibinfo
  {pages} {1073} (\bibinfo {year} {2010})}\BibitemShut {NoStop}%
\bibitem [{\citenamefont {Bhushan}\ \emph {et~al.}(1995)\citenamefont
  {Bhushan}, \citenamefont {Israelachvili},\ and\ \citenamefont
  {Landman}}]{Bhu95}%
  \BibitemOpen
  \bibfield  {author} {\bibinfo {author} {\bibfnamefont {B.}~\bibnamefont
  {Bhushan}}, \bibinfo {author} {\bibfnamefont {J.~N.}\ \bibnamefont
  {Israelachvili}}, \ and\ \bibinfo {author} {\bibfnamefont {U.}~\bibnamefont
  {Landman}},\ }\href@noop {} {\bibfield  {journal} {\bibinfo  {journal}
  {Nature}\ }\textbf {\bibinfo {volume} {374}},\ \bibinfo {pages} {607}
  (\bibinfo {year} {1995})}\BibitemShut {NoStop}%
\bibitem [{\citenamefont {Cheng}\ \emph {et~al.}(2012)\citenamefont {Cheng},
  \citenamefont {Hu},\ and\ \citenamefont {Hu}}]{Che12}%
  \BibitemOpen
  \bibfield  {author} {\bibinfo {author} {\bibfnamefont {H.~F.}\ \bibnamefont
  {Cheng}}, \bibinfo {author} {\bibfnamefont {E.~D.}\ \bibnamefont {Hu}}, \
  and\ \bibinfo {author} {\bibfnamefont {Y.~A.}\ \bibnamefont {Hu}},\
  }\href@noop {} {\bibfield  {journal} {\bibinfo  {journal} {Journal of
  Contaminant Hydrology}\ }\textbf {\bibinfo {volume} {129}},\ \bibinfo {pages}
  {80} (\bibinfo {year} {2012})}\BibitemShut {NoStop}%
\bibitem [{\citenamefont {Finney}(1996)}]{Fin96}%
  \BibitemOpen
  \bibfield  {author} {\bibinfo {author} {\bibfnamefont {J.~L.}\ \bibnamefont
  {Finney}},\ }\href@noop {} {\bibfield  {journal} {\bibinfo  {journal}
  {Faraday Discussions}\ }\textbf {\bibinfo {volume} {103}},\ \bibinfo {pages}
  {1} (\bibinfo {year} {1996})}\BibitemShut {NoStop}%
\bibitem [{\citenamefont {Prakash}\ \emph {et~al.}(2012)\citenamefont
  {Prakash}, \citenamefont {Pinti},\ and\ \citenamefont {Bhushan}}]{Pra12}%
  \BibitemOpen
  \bibfield  {author} {\bibinfo {author} {\bibfnamefont {S.}~\bibnamefont
  {Prakash}}, \bibinfo {author} {\bibfnamefont {M.}~\bibnamefont {Pinti}}, \
  and\ \bibinfo {author} {\bibfnamefont {B.}~\bibnamefont {Bhushan}},\
  }\href@noop {} {\bibfield  {journal} {\bibinfo  {journal} {Philosophical
  Transactions of the Royal Society a-Mathematical Physical and Engineering
  Sciences}\ }\textbf {\bibinfo {volume} {370}},\ \bibinfo {pages} {2269}
  (\bibinfo {year} {2012})}\BibitemShut {NoStop}%
\bibitem [{\citenamefont {Klein}\ and\ \citenamefont
  {Kumacheva}(1998)}]{Kle98}%
  \BibitemOpen
  \bibfield  {author} {\bibinfo {author} {\bibfnamefont {J.}~\bibnamefont
  {Klein}}\ and\ \bibinfo {author} {\bibfnamefont {E.}~\bibnamefont
  {Kumacheva}},\ }\href@noop {} {\bibfield  {journal} {\bibinfo  {journal}
  {Journal of Chemical Physics}\ }\textbf {\bibinfo {volume} {108}},\ \bibinfo
  {pages} {6996} (\bibinfo {year} {1998})}\BibitemShut {NoStop}%
\bibitem [{\citenamefont {Antognozzi}\ \emph {et~al.}(2001)\citenamefont
  {Antognozzi}, \citenamefont {Humphris},\ and\ \citenamefont {Miles}}]{Ant01}%
  \BibitemOpen
  \bibfield  {author} {\bibinfo {author} {\bibfnamefont {M.}~\bibnamefont
  {Antognozzi}}, \bibinfo {author} {\bibfnamefont {A.~D.~L.}\ \bibnamefont
  {Humphris}}, \ and\ \bibinfo {author} {\bibfnamefont {M.~J.}\ \bibnamefont
  {Miles}},\ }\href@noop {} {\bibfield  {journal} {\bibinfo  {journal} {Applied
  Physics Letters}\ }\textbf {\bibinfo {volume} {78}},\ \bibinfo {pages} {300}
  (\bibinfo {year} {2001})}\BibitemShut {NoStop}%
\bibitem [{\citenamefont {Demirel}\ and\ \citenamefont
  {Granick}(2001)}]{Dem01}%
  \BibitemOpen
  \bibfield  {author} {\bibinfo {author} {\bibfnamefont {A.~L.}\ \bibnamefont
  {Demirel}}\ and\ \bibinfo {author} {\bibfnamefont {S.}~\bibnamefont
  {Granick}},\ }\href@noop {} {\bibfield  {journal} {\bibinfo  {journal}
  {Journal of Chemical Physics}\ }\textbf {\bibinfo {volume} {115}},\ \bibinfo
  {pages} {1498} (\bibinfo {year} {2001})}\BibitemShut {NoStop}%
\bibitem [{\citenamefont {Raviv}\ and\ \citenamefont {Klein}(2002)}]{Rav02}%
  \BibitemOpen
  \bibfield  {author} {\bibinfo {author} {\bibfnamefont {U.}~\bibnamefont
  {Raviv}}\ and\ \bibinfo {author} {\bibfnamefont {J.}~\bibnamefont {Klein}},\
  }\href@noop {} {\bibfield  {journal} {\bibinfo  {journal} {Science}\ }\textbf
  {\bibinfo {volume} {297}},\ \bibinfo {pages} {1540} (\bibinfo {year}
  {2002})}\BibitemShut {NoStop}%
\bibitem [{\citenamefont {Becker}\ and\ \citenamefont {Mugele}(2003)}]{Bec03}%
  \BibitemOpen
  \bibfield  {author} {\bibinfo {author} {\bibfnamefont {T.}~\bibnamefont
  {Becker}}\ and\ \bibinfo {author} {\bibfnamefont {F.}~\bibnamefont
  {Mugele}},\ }\href@noop {} {\bibfield  {journal} {\bibinfo  {journal}
  {Physical Review Letters}\ }\textbf {\bibinfo {volume} {91}},\ \bibinfo
  {pages} {166104 (1} (\bibinfo {year} {2003})}\BibitemShut {NoStop}%
\bibitem [{\citenamefont {Li}\ \emph {et~al.}(2007)\citenamefont {Li},
  \citenamefont {Gao}, \citenamefont {Szoszkiewicz}, \citenamefont {Landman},\
  and\ \citenamefont {Riedo}}]{Li07}%
  \BibitemOpen
  \bibfield  {author} {\bibinfo {author} {\bibfnamefont {T.~D.}\ \bibnamefont
  {Li}}, \bibinfo {author} {\bibfnamefont {J.~P.}\ \bibnamefont {Gao}},
  \bibinfo {author} {\bibfnamefont {R.}~\bibnamefont {Szoszkiewicz}}, \bibinfo
  {author} {\bibfnamefont {U.}~\bibnamefont {Landman}}, \ and\ \bibinfo
  {author} {\bibfnamefont {E.}~\bibnamefont {Riedo}},\ }\href@noop {}
  {\bibfield  {journal} {\bibinfo  {journal} {Physical Review B}\ }\textbf
  {\bibinfo {volume} {75}},\ \bibinfo {pages} {115415} (\bibinfo {year}
  {2007})}\BibitemShut {NoStop}%
\bibitem [{\citenamefont {Matei}\ \emph {et~al.}(2008)\citenamefont {Matei},
  \citenamefont {Jeffery}, \citenamefont {Patil}, \citenamefont {Khan},
  \citenamefont {Pantea}, \citenamefont {Pethica},\ and\ \citenamefont
  {Hoffmann}}]{Mat08}%
  \BibitemOpen
  \bibfield  {author} {\bibinfo {author} {\bibfnamefont {G.}~\bibnamefont
  {Matei}}, \bibinfo {author} {\bibfnamefont {S.}~\bibnamefont {Jeffery}},
  \bibinfo {author} {\bibfnamefont {S.}~\bibnamefont {Patil}}, \bibinfo
  {author} {\bibfnamefont {S.~H.}\ \bibnamefont {Khan}}, \bibinfo {author}
  {\bibfnamefont {M.}~\bibnamefont {Pantea}}, \bibinfo {author} {\bibfnamefont
  {J.~B.}\ \bibnamefont {Pethica}}, \ and\ \bibinfo {author} {\bibfnamefont
  {P.~M.}\ \bibnamefont {Hoffmann}},\ }\href@noop {} {\bibfield  {journal}
  {\bibinfo  {journal} {Review of Scientific Instruments}\ }\textbf {\bibinfo
  {volume} {79}},\ \bibinfo {pages} {023706} (\bibinfo {year}
  {2008})}\BibitemShut {NoStop}%
\bibitem [{\citenamefont {Kaggwa}\ \emph {et~al.}(2008)\citenamefont {Kaggwa},
  \citenamefont {Kilpatrick}, \citenamefont {Sader},\ and\ \citenamefont
  {Jarvis}}]{Kag08}%
  \BibitemOpen
  \bibfield  {author} {\bibinfo {author} {\bibfnamefont {G.~B.}\ \bibnamefont
  {Kaggwa}}, \bibinfo {author} {\bibfnamefont {J.~I.}\ \bibnamefont
  {Kilpatrick}}, \bibinfo {author} {\bibfnamefont {J.~E.}\ \bibnamefont
  {Sader}}, \ and\ \bibinfo {author} {\bibfnamefont {S.~P.}\ \bibnamefont
  {Jarvis}},\ }\href@noop {} {\bibfield  {journal} {\bibinfo  {journal}
  {Applied Physics Letters}\ }\textbf {\bibinfo {volume} {93}},\ \bibinfo
  {pages} {3} (\bibinfo {year} {2008})}\BibitemShut {NoStop}%
\bibitem [{\citenamefont {Hofbauer}\ \emph {et~al.}(2009)\citenamefont
  {Hofbauer}, \citenamefont {Ho}, \citenamefont {Hairulnizam}, \citenamefont
  {Gosvami},\ and\ \citenamefont {O'Shea}}]{Hof09}%
  \BibitemOpen
  \bibfield  {author} {\bibinfo {author} {\bibfnamefont {W.}~\bibnamefont
  {Hofbauer}}, \bibinfo {author} {\bibfnamefont {R.~J.}\ \bibnamefont {Ho}},
  \bibinfo {author} {\bibfnamefont {R.}~\bibnamefont {Hairulnizam}}, \bibinfo
  {author} {\bibfnamefont {N.~N.}\ \bibnamefont {Gosvami}}, \ and\ \bibinfo
  {author} {\bibfnamefont {S.~J.}\ \bibnamefont {O'Shea}},\ }\href@noop {}
  {\bibfield  {journal} {\bibinfo  {journal} {Physical Review B}\ }\textbf
  {\bibinfo {volume} {80}},\ \bibinfo {pages} {134104 (1} (\bibinfo {year}
  {2009})}\BibitemShut {NoStop}%
\bibitem [{\citenamefont {Li}\ and\ \citenamefont {Riedo}(2008)}]{Li08}%
  \BibitemOpen
  \bibfield  {author} {\bibinfo {author} {\bibfnamefont {T.~D.}\ \bibnamefont
  {Li}}\ and\ \bibinfo {author} {\bibfnamefont {E.}~\bibnamefont {Riedo}},\
  }\href@noop {} {\bibfield  {journal} {\bibinfo  {journal} {Physical Review
  Letters}\ }\textbf {\bibinfo {volume} {100}},\ \bibinfo {pages} {106102}
  (\bibinfo {year} {2008})}\BibitemShut {NoStop}%
\bibitem [{\citenamefont {Patil}\ \emph {et~al.}(2006)\citenamefont {Patil},
  \citenamefont {Matei}, \citenamefont {Oral},\ and\ \citenamefont
  {Hoffmann}}]{Pat06}%
  \BibitemOpen
  \bibfield  {author} {\bibinfo {author} {\bibfnamefont {S.}~\bibnamefont
  {Patil}}, \bibinfo {author} {\bibfnamefont {G.}~\bibnamefont {Matei}},
  \bibinfo {author} {\bibfnamefont {A.}~\bibnamefont {Oral}}, \ and\ \bibinfo
  {author} {\bibfnamefont {P.~M.}\ \bibnamefont {Hoffmann}},\ }\href@noop {}
  {\bibfield  {journal} {\bibinfo  {journal} {Langmuir}\ }\textbf {\bibinfo
  {volume} {22}},\ \bibinfo {pages} {6485} (\bibinfo {year}
  {2006})}\BibitemShut {NoStop}%
\bibitem [{\citenamefont {Khan}\ \emph {et~al.}(2010)\citenamefont {Khan},
  \citenamefont {Matei}, \citenamefont {Patil},\ and\ \citenamefont
  {Hoffmann}}]{Kha10}%
  \BibitemOpen
  \bibfield  {author} {\bibinfo {author} {\bibfnamefont {S.~H.}\ \bibnamefont
  {Khan}}, \bibinfo {author} {\bibfnamefont {G.}~\bibnamefont {Matei}},
  \bibinfo {author} {\bibfnamefont {S.}~\bibnamefont {Patil}}, \ and\ \bibinfo
  {author} {\bibfnamefont {P.~M.}\ \bibnamefont {Hoffmann}},\ }\href@noop {}
  {\bibfield  {journal} {\bibinfo  {journal} {Physical Review Letters}\
  }\textbf {\bibinfo {volume} {105}},\ \bibinfo {pages} {106101} (\bibinfo
  {year} {2010})}\BibitemShut {NoStop}%
\bibitem [{\citenamefont {de~Beer}\ \emph {et~al.}(2010)\citenamefont
  {de~Beer}, \citenamefont {van~den Ende},\ and\ \citenamefont
  {Mugele}}]{deB10}%
  \BibitemOpen
  \bibfield  {author} {\bibinfo {author} {\bibfnamefont {S.}~\bibnamefont
  {de~Beer}}, \bibinfo {author} {\bibfnamefont {D.}~\bibnamefont {van~den
  Ende}}, \ and\ \bibinfo {author} {\bibfnamefont {F.}~\bibnamefont {Mugele}},\
  }\href@noop {} {\bibfield  {journal} {\bibinfo  {journal} {Nanotechnology}\
  }\textbf {\bibinfo {volume} {21}} (\bibinfo {year} {2010})}\BibitemShut
  {NoStop}%
\bibitem [{\citenamefont {Zhu}\ and\ \citenamefont {Granick}(2003)}]{Zhu03}%
  \BibitemOpen
  \bibfield  {author} {\bibinfo {author} {\bibfnamefont {Y.}~\bibnamefont
  {Zhu}}\ and\ \bibinfo {author} {\bibfnamefont {S.}~\bibnamefont {Granick}},\
  }\href@noop {} {\bibfield  {journal} {\bibinfo  {journal} {Langmuir}\
  }\textbf {\bibinfo {volume} {19}},\ \bibinfo {pages} {8148} (\bibinfo {year}
  {2003})}\BibitemShut {NoStop}%
\bibitem [{\citenamefont {Bureau}(2010)}]{Bur10}%
  \BibitemOpen
  \bibfield  {author} {\bibinfo {author} {\bibfnamefont {L.}~\bibnamefont
  {Bureau}},\ }\href@noop {} {\bibfield  {journal} {\bibinfo  {journal}
  {Physical Review Letters}\ }\textbf {\bibinfo {volume} {104}} (\bibinfo
  {year} {2010})}\BibitemShut {NoStop}%
\bibitem [{\citenamefont {O'Shea}\ and\ \citenamefont {Welland}(1998)}]{Osh98}%
  \BibitemOpen
  \bibfield  {author} {\bibinfo {author} {\bibfnamefont {S.~J.}\ \bibnamefont
  {O'Shea}}\ and\ \bibinfo {author} {\bibfnamefont {M.~E.}\ \bibnamefont
  {Welland}},\ }\href@noop {} {\bibfield  {journal} {\bibinfo  {journal}
  {Langmuir}\ }\textbf {\bibinfo {volume} {14}},\ \bibinfo {pages} {4186}
  (\bibinfo {year} {1998})}\BibitemShut {NoStop}%
\bibitem [{\citenamefont {Derjaguin}(1934)}]{Der34}%
  \BibitemOpen
  \bibfield  {author} {\bibinfo {author} {\bibfnamefont {B.~V.}\ \bibnamefont
  {Derjaguin}},\ }\href@noop {} {\bibfield  {journal} {\bibinfo  {journal}
  {Kolloid Zeitschrift}\ }\textbf {\bibinfo {volume} {69}},\ \bibinfo {pages}
  {155} (\bibinfo {year} {1934})}\BibitemShut {NoStop}%
\bibitem [{\citenamefont {Williams}(2005)}]{Wil05}%
  \BibitemOpen
  \bibfield  {author} {\bibinfo {author} {\bibfnamefont {J.~A.}\ \bibnamefont
  {Williams}},\ }\href@noop {} {\emph {\bibinfo {title} {Engineering
  tribology}}}\ (\bibinfo  {publisher} {Cambridge University Press},\ \bibinfo
  {address} {New York},\ \bibinfo {year} {2005})\BibitemShut {NoStop}%
\bibitem [{\citenamefont {Yu}\ \emph {et~al.}(1999)\citenamefont {Yu},
  \citenamefont {Richter}, \citenamefont {Datta}, \citenamefont {Durbin},\ and\
  \citenamefont {Dutta}}]{Yu99}%
  \BibitemOpen
  \bibfield  {author} {\bibinfo {author} {\bibfnamefont {C.~J.}\ \bibnamefont
  {Yu}}, \bibinfo {author} {\bibfnamefont {A.~G.}\ \bibnamefont {Richter}},
  \bibinfo {author} {\bibfnamefont {A.}~\bibnamefont {Datta}}, \bibinfo
  {author} {\bibfnamefont {M.~K.}\ \bibnamefont {Durbin}}, \ and\ \bibinfo
  {author} {\bibfnamefont {P.}~\bibnamefont {Dutta}},\ }\href@noop {}
  {\bibfield  {journal} {\bibinfo  {journal} {Physical Review Letters}\
  }\textbf {\bibinfo {volume} {82}},\ \bibinfo {pages} {2326} (\bibinfo {year}
  {1999})}\BibitemShut {NoStop}%
\bibitem [{\citenamefont {Patil}\ \emph {et~al.}(2007)\citenamefont {Patil},
  \citenamefont {Matei}, \citenamefont {Grabowski}, \citenamefont {Hoffmann},\
  and\ \citenamefont {Mukhopadhyay}}]{Pat07}%
  \BibitemOpen
  \bibfield  {author} {\bibinfo {author} {\bibfnamefont {S.}~\bibnamefont
  {Patil}}, \bibinfo {author} {\bibfnamefont {G.}~\bibnamefont {Matei}},
  \bibinfo {author} {\bibfnamefont {C.~A.}\ \bibnamefont {Grabowski}}, \bibinfo
  {author} {\bibfnamefont {P.~M.}\ \bibnamefont {Hoffmann}}, \ and\ \bibinfo
  {author} {\bibfnamefont {A.}~\bibnamefont {Mukhopadhyay}},\ }\href@noop {}
  {\bibfield  {journal} {\bibinfo  {journal} {Langmuir}\ }\textbf {\bibinfo
  {volume} {23}},\ \bibinfo {pages} {4988} (\bibinfo {year}
  {2007})}\BibitemShut {NoStop}%
\bibitem [{\citenamefont {Patil}\ \emph {et~al.}(2005)\citenamefont {Patil},
  \citenamefont {Matei}, \citenamefont {Dong}, \citenamefont {Hoffmann},
  \citenamefont {Karakose},\ and\ \citenamefont {Oral}}]{Pat05}%
  \BibitemOpen
  \bibfield  {author} {\bibinfo {author} {\bibfnamefont {S.}~\bibnamefont
  {Patil}}, \bibinfo {author} {\bibfnamefont {G.}~\bibnamefont {Matei}},
  \bibinfo {author} {\bibfnamefont {H.}~\bibnamefont {Dong}}, \bibinfo {author}
  {\bibfnamefont {P.~M.}\ \bibnamefont {Hoffmann}}, \bibinfo {author}
  {\bibfnamefont {M.}~\bibnamefont {Karakose}}, \ and\ \bibinfo {author}
  {\bibfnamefont {A.}~\bibnamefont {Oral}},\ }\href@noop {} {\bibfield
  {journal} {\bibinfo  {journal} {Review of Scientific Instruments}\ }\textbf
  {\bibinfo {volume} {76}},\ \bibinfo {pages} {103705} (\bibinfo {year}
  {2005})}\BibitemShut {NoStop}%
\bibitem [{\citenamefont {Lim}\ \emph {et~al.}(2008)\citenamefont {Lim},
  \citenamefont {Wee},\ and\ \citenamefont {O'Shea}}]{Lim08}%
  \BibitemOpen
  \bibfield  {author} {\bibinfo {author} {\bibfnamefont {L.~T.~W.}\
  \bibnamefont {Lim}}, \bibinfo {author} {\bibfnamefont {A.~T.~S.}\
  \bibnamefont {Wee}}, \ and\ \bibinfo {author} {\bibfnamefont {S.~J.}\
  \bibnamefont {O'Shea}},\ }\href@noop {} {\bibfield  {journal} {\bibinfo
  {journal} {Langmuir}\ }\textbf {\bibinfo {volume} {24}},\ \bibinfo {pages}
  {2271} (\bibinfo {year} {2008})}\BibitemShut {NoStop}%
\bibitem [{\citenamefont {Leroy}\ and\ \citenamefont {Charlaix}(2011)}]{Ler11}%
  \BibitemOpen
  \bibfield  {author} {\bibinfo {author} {\bibfnamefont {S.}~\bibnamefont
  {Leroy}}\ and\ \bibinfo {author} {\bibfnamefont {E.}~\bibnamefont
  {Charlaix}},\ }\href@noop {} {\bibfield  {journal} {\bibinfo  {journal}
  {Journal of Fluid Mechanics}\ }\textbf {\bibinfo {volume} {674}},\ \bibinfo
  {pages} {389} (\bibinfo {year} {2011})}\BibitemShut {NoStop}%
\bibitem [{\citenamefont {Hu}\ \emph {et~al.}(1991)\citenamefont {Hu},
  \citenamefont {Carson},\ and\ \citenamefont {Granick}}]{Hu91}%
  \BibitemOpen
  \bibfield  {author} {\bibinfo {author} {\bibfnamefont {H.~W.}\ \bibnamefont
  {Hu}}, \bibinfo {author} {\bibfnamefont {G.~A.}\ \bibnamefont {Carson}}, \
  and\ \bibinfo {author} {\bibfnamefont {S.}~\bibnamefont {Granick}},\
  }\href@noop {} {\bibfield  {journal} {\bibinfo  {journal} {Physical Review
  Letters}\ }\textbf {\bibinfo {volume} {66}},\ \bibinfo {pages} {2758}
  (\bibinfo {year} {1991})}\BibitemShut {NoStop}%
\bibitem [{\citenamefont {Bird}\ and\ \citenamefont {Carreau}(1968)}]{Bir68}%
  \BibitemOpen
  \bibfield  {author} {\bibinfo {author} {\bibfnamefont {R.~B.}\ \bibnamefont
  {Bird}}\ and\ \bibinfo {author} {\bibfnamefont {P.~J.}\ \bibnamefont
  {Carreau}},\ }\href@noop {} {\bibfield  {journal} {\bibinfo  {journal}
  {Chemical Engineering Science}\ }\textbf {\bibinfo {volume} {23}},\ \bibinfo
  {pages} {427} (\bibinfo {year} {1968})}\BibitemShut {NoStop}%
\bibitem [{\citenamefont {Abbott}\ \emph {et~al.}(1961)\citenamefont {Abbott},
  \citenamefont {Wright}, \citenamefont {Goldschmidt}, \citenamefont
  {Stewart},\ and\ \citenamefont {Bolt}}]{Abb61}%
  \BibitemOpen
  \bibfield  {author} {\bibinfo {author} {\bibfnamefont {A.~D.}\ \bibnamefont
  {Abbott}}, \bibinfo {author} {\bibfnamefont {J.~R.}\ \bibnamefont {Wright}},
  \bibinfo {author} {\bibfnamefont {A.}~\bibnamefont {Goldschmidt}}, \bibinfo
  {author} {\bibfnamefont {W.~T.}\ \bibnamefont {Stewart}}, \ and\ \bibinfo
  {author} {\bibfnamefont {R.~O.}\ \bibnamefont {Bolt}},\ }\href@noop {}
  {\bibfield  {journal} {\bibinfo  {journal} {J. Chem. Eng. Data}\ }\textbf
  {\bibinfo {volume} {6}},\ \bibinfo {pages} {437} (\bibinfo {year}
  {1961})}\BibitemShut {NoStop}%
\bibitem [{\citenamefont {Hoffmann}\ \emph {et~al.}(2001)\citenamefont
  {Hoffmann}, \citenamefont {Oral}, \citenamefont {Grimble}, \citenamefont
  {Ozer}, \citenamefont {Jeffery},\ and\ \citenamefont {Pethica}}]{Hof01}%
  \BibitemOpen
  \bibfield  {author} {\bibinfo {author} {\bibfnamefont {P.~M.}\ \bibnamefont
  {Hoffmann}}, \bibinfo {author} {\bibfnamefont {A.}~\bibnamefont {Oral}},
  \bibinfo {author} {\bibfnamefont {R.~A.}\ \bibnamefont {Grimble}}, \bibinfo
  {author} {\bibfnamefont {H.~O.}\ \bibnamefont {Ozer}}, \bibinfo {author}
  {\bibfnamefont {S.}~\bibnamefont {Jeffery}}, \ and\ \bibinfo {author}
  {\bibfnamefont {J.~B.}\ \bibnamefont {Pethica}},\ }\href@noop {} {\bibfield
  {journal} {\bibinfo  {journal} {Proceedings of the Royal Society of London
  Series a-Mathematical Physical and Engineering Sciences}\ }\textbf {\bibinfo
  {volume} {457}},\ \bibinfo {pages} {1161} (\bibinfo {year}
  {2001})}\BibitemShut {NoStop}%
\end{thebibliography}%

\end{document}